\documentclass[doublecol]{epl2}

\input{tcilatex}

\title{Magnetization reversal and nonexponential relaxation via instabilities of internal spin waves in nanomagnets}
\shorttitle{Spin-wave relaxation in magnetic nanoparticles}

\author{D. A. Garanin\inst{1} \and H. Kachkachi\inst{2} \and L. Reynaud\inst{2}}
\shortauthor{Garanin \etal}

\institute{
  \inst{1} Physics Department, Lehman College, City University of New York, 250 Bedford Park Boulevard West, Bronx, New York 10468-1589, U.S.A.\\
  \inst{2} Groupe d'Etude de la Mati\`ere Condens\'ee, CNRS UMR8634, Universit\'e de Versailles St. Quentin, 45 av. des Etats-Unis, 78035 Versailles, France
}
\abstract{
A magnetic particle with atomic spins ordered in an unstable direction is an
example of a false vacuum that decays via excitation of internal spin waves.
Coupled evolution of the particle's magnetization (or the vacuum state) and
spin waves, considered in the time-dependent vacuum frame, leads to a
peculiar relaxation that is very fast at the beginning but slows down to a
nonexponential long tail at the end. The two main scenarios are linear and
exponential spin-wave instabilities. For the former, the longitudinal and
transverse relaxation rates have been obtained analytically. Numerical
simulations show that the particle's magnetization strongly decreases in the
middle of reversal and then recovers.
}
\begin{document}
\maketitle
\section{Introduction}
\label{sec-intro}
Magnetization reversal in magnetic particles requires transfer of their
excessive Zeeman and/or anisotropy energy into a heat bath. Magnetic
particles are usually considered as \emph{single magnetic moments} relaxing
to external heat reservoirs such as phonons, conduction electrons, etc. Very
popular is a semi-phenomenological description in terms of the
Landau-Lifshitz damping and Langevin noise field (see, e.g., Refs. \cite{bro63,gar97prb,garfes04prb}). The resulting relaxation is exponential or
close to it.

On the other hand, for larger single-domain particles that can accommodate
internal spin waves, the latter can serve as a heat bath and thus give rise
to the particle's relaxation and magnetization switching \cite{suh98,safber01prb,saf04jap}. Internal relaxation processes should be much faster than those that operate via nonmagnetic degrees of freedom, with important implications for engineering, determining response rates of magnetic elements \cite{schgerkossil05apl}.

Spin-wave (SW) dynamics in magnetic particles whose magnetization changes in
time is a new and challenging topic. Of course, small deviations from
equilibrium, including the case of Suhl instabilities \cite
{suh57,suh98,dobvic03prl,dobvic04jap,lounalkle05prbrc}, can still be
described with the help of a nonlinear SW theory built around the ground
state. In general, however, one has to redefine the \emph{spin-wave vacuum}
by considering SW dynamics in the frame of the time dependent particle's
magnetization \cite{kas06prl}. Earlier, a similar approach was suggested for
the thermodynamics of low-dimensional magnetic systems \cite{hasnie93} and
that of magnetic particles \cite{kacgar01epjb}.

The false-vacuum initial condition can be created by biasing the
system with a magnetic field that leads to the disappearance of the
metastable energy minimum \cite{safber01prb}. Another method is the
precessional switching \cite{backetal99science} that allows to rotate the
magnetization by ultrafast magnetic field pulses as short as 2 picoseconds
with a field amplitude as small as 184 kiloamperes per meter.

Up to date, the coupled motion of the particle's magnetization (the vacuum) and
internal spin waves is not yet completely understood analytically, and
different scenarios of relaxation are not yet classified. In addition, the relaxation from
homogeneous states far from equilibrium has not yet been traced numerically
over large enough times for large enough particles to see all stages of
relaxation.

This Letter presents analytical and numerical work showing that the
interplay between the global and local modes (the magnetization $\mathbf{m}$
and spin waves) leads to an initially very fast relaxation due to SW
instabilities which slows down to a long nonexponential tail in its
concluding stage. Such a kind of relaxation seems to be unknown in magnetism.

The numerically observed magnetization switching is almost complete, and it is
typically accompanied by a strong reduction of the particle's magnetization $m\equiv |\mathbf{m}|$ as the spin waves are excited and the system deviates from the initial false-vacuum state. Subsequently $m$ recovers to slightly less than the initial value $m\cong 1$ as the excited SWs dissipate into the small-amplitude short-wavelength excitations and the system approaches the
true vacuum state with some thermal disordering. This longitudinal relaxation resembles that described by the Landau-Lifshitz-Bloch (LLB) equation \cite{gar97prb,garfes04prb}.

One can identify two main scenarios of the relaxation via internal spin
waves. These are the \emph{linear }and \emph{exponential} instabilities,
characterized by the corresponding time dependences $\Delta m(t)\equiv
m(t)-m(0)$ at short times.

The linear instability is a one-magnon process of the decay of the unstable
(false) SW vacuum into the \emph{lower-lying} SW quasicontinuum. In the
standard SW theory operating near the energy minimum, the one-magnon process is
of course absent. This mechanism occurs in all cases of slightly noncollinear
magnetization that include dipole-dipole interaction (DDI), surface
anisotropy, and random anisotropy. The latter will be used below for
analytical and numerical work as the most transparent case. As the magnetization
gradually rotates to its equilibrium direction, the false SW vacuum becomes
closer to the true vacuum and thus less unstable. This leads to a slowing
down of the relaxation.

The \emph{exponential instability} in the case of non-circular magnetization
precession can be realized in models with biaxial anisotropy, uniaxial
anisotropy with transverse magnetic field, DDI, and many others. Physically,
this mechanism is a kind of Suhl instability \cite{suh57,suh98,dobvic03prl}.
However, in the global-magnetization frame it finds a much simpler
analytical description \cite{kas06prl} that is not limited to small
deviations from equilibrium. In this case too, numerical simulations
show a long magnetization tail, after a very agile initial relaxation.

\section{Spin model and relaxation formalism}
\label{sec-hamiltonian}
We consider a classical spin model ($|\mathbf{s}_{i}|=1$)
\begin{equation}\label{Ham}
\mathcal{H}=\mathcal{H}_{A}-\mathbf{h}\cdot \sum_{i}\mathbf{s}_{i}-\frac{1}{2}\sum_{ij}J_{ij}\mathbf{s}_{i}\cdot \mathbf{s}_{j},
\end{equation}
where $\mathbf{h}=g\mu _{B}\mathbf{H}$ is the magnetic field in energy units, $J_{ij}$ is the exchange, and $\mathcal{H}_{A}$ is the anisotropy energy. In the present work, we consider two models of anisotropy. The first is that of random uniaxial anisotropy with the energy
\begin{equation}\label{RADef}
\mathcal{H}_{A} = -D_{R}\sum_{i}\left(\mathbf{u}_{i}\cdot\mathbf{s}_{i}\right)^{2},\qquad D_{R} > 0,
\end{equation}
where $\mathbf{u}_{i}$ is a random unit vector. The latter does not break the global particle's isotropy, and its role is merely to provide amplitudes for SW processes that lead to non-conservation of $\mathbf{m}\mathbf{\cdot h}$ and thus to reversal.
The second model is that of oriented uniaxial anisotropy, i.e., all spins have their easy axes lying in the same reference $z$ direction, and thus
\begin{equation}\label{OADef}
\mathcal{H}_{A} = -D\sum_{i}s_{iz}^{2},\qquad D > 0.
\end{equation}

We define the particle's magnetization as
\begin{equation}\label{mDef}
\mathbf{m}=\frac{1}{\mathcal{N}}\sum_{i}\mathbf{s}_{i},
\end{equation}
where $\mathcal{N}$ is the total number of spins.

The atomic spins obey the Larmor equation
\begin{equation}\label{LLEqi}
\mathbf{\dot{s}}_{i}=\left[ \mathbf{s}_{i}\times \mathbf{\Omega }_{i}\right]
,\qquad \hbar \mathbf{\Omega }_{i}=-\partial \mathcal{H/}\partial \mathbf{s}%
_{i}.
\end{equation}
We do not include any \emph{ad hoc} relaxation terms, so that all relaxation that we observe is due to the intrinsic nonlinearities in Eq. (\ref{Ham}).

Now, we present the formalism of relaxation via internal spin waves using the Hamiltonian (\ref{Ham}) with $\mathcal{H}_{A}$ being given by either (\ref{RADef}) or (\ref{OADef}).
For convenience, we keep both these terms in general calculations.
The effective field $\mathbf{\Omega }_{i}$ in Eq.\ (\ref{LLEqi}) in vector-tensor notations becomes
\begin{equation}\label{Omegai}
\hbar\mathbf{\Omega}_{i}=\mathbf{h}+2(\mathbf{D}+\mathbf{g}_{i})\cdot\mathbf{s}_{i}+\sum_{j}J_{ij}\mathbf{s}_{j}
\end{equation}
where $(\mathbf{D})_{\alpha \beta } = D\delta _{\alpha \beta z}$ and $(\mathbf{g}_i)_{\alpha \beta}
=D_{R}u_{i,\alpha }u_{i,\beta }$, with $D$ and $D_R$ being defined in Eqs. (\ref{RADef}) and (\ref{OADef}), respectively.

To construct a spin-wave theory around the false vacuum, one has to split $\mathbf{s}_{i}$ into two parts describing respectively the vacuum mode [the particle's magnetization $\mathbf{m}$ of Eq.
(\ref{mDef})] and the internal spin waves, that is
\begin{equation}
\mathbf{s}_{i}=\mathbf{m}+\mathbf{\psi }_{i},\qquad \sum_{i}\mathbf{\psi }%
_{i}=0.  \label{psiDef}
\end{equation}
Whereas in the standard SW theory $\mathbf{m}$ is a constant corresponding to the ground-state orientation, here it is treated as a time-dependent variable. Since the atomic spins are subject to the chiral constraint $\mathbf{s}%
_{i}^{2}=1,$ one can use \cite{kas06prl}
$
\mathbf{m=n}\sqrt{1-\psi _{i}^{2}}$ with $ \mathbf{n}\cdot \mathbf{\psi }%
_{i}=0,
$
where $\mathbf{n}$ is a unit vector. Although this reduces the number of the $\mathbf{\psi }_{i}$ components in the frame related to the particle's magnetization $\mathbf{m}$ to only two, the formalism becomes much more cumbersome, whereas the final results are not affected. Thus we decided not to use the chiral constraint explicitly in our presentation. Of course, properly written equations must satisfy this constraint that can be used to check them. The equation of motion for $\mathbf{m}$ following from Eq.\ (\ref{LLEqi}) has the form
\begin{equation}
\hbar \mathbf{\dot{m}}=\left[ \mathbf{m\times }\left( \mathbf{h}+2\mathbf{D}\cdot\mathbf{m}\right) \right] +\mathbf{R,}
\label{mEq0}
\end{equation}
where $\mathbf{R}$ couples the dynamics of $\mathbf{m}$ to spin waves
\begin{eqnarray}
\mathbf{R} &=&\frac{1}{\mathcal{N}}\sum_{i}\left\{ \left[ \mathbf{m\times }
(2\mathbf{g}_{i}\cdot\mathbf{\psi }_{i})\right] \right.  \nonumber
\\
&&\quad +\left. \left[ \mathbf{\psi }_{i}\times \left( 2
\mathbf{D}\cdot\mathbf{\psi }_{i}+2\mathbf{g}_{i}\cdot\left(
\mathbf{m+\psi }_{i}\right) \right) \right] \right\} .  \label{RDef}
\end{eqnarray}
In turn, the equation of motion for $\mathbf{\psi }_{i}$ has the form
\begin{equation}
\hbar \mathbf{\dot{\psi}}_{i}=\left[ \mathbf{m\times }(2
\mathbf{g}_{i}\cdot\mathbf{m})\right] +\mathbf{A}_{i}^{(1)}+\mathbf{A}_{i}^{(2)},
\label{psiiEq}
\end{equation}
where $\mathbf{A}_{i}^{(1)}$ is linear in $\mathbf{\psi }_{i}$
\begin{eqnarray}
\mathbf{A}_{i}^{(1)} &=&\left[ \mathbf{\psi }_{i}\times \left( \mathbf{h}+%
2\mathbf{D}\cdot\mathbf{m}+J_{0}\mathbf{m}\right) \right]
\nonumber \\
&&+\left[ \mathbf{m}\times \left( 2\mathbf{D}\cdot\mathbf{%
\psi }_{i}+\sum_{j}J_{ij}\mathbf{\psi }_{j}\right) \right]  \label{A1iDef}
\end{eqnarray}
and $\mathbf{A}_{i}^{(2)}$ contains terms of order $\psi^2$ and $\psi g.$ The first term in Eq.\ (\ref{psiiEq}), that causes the spin non-collinearity, is responsible for the linear SW instability via direct generation of spin waves. We remark in passing that the same effect is produced by e.g., surface anisotropy and dipole-dipole interactions.
The terms containing $\mathbf{D}$ in $\mathbf{A}_{i}^{(1)}$ can be responsible for the exponential SW instability. Since these terms are linear in $\mathbf{\psi }_{i}$, this instability requires some SWs in the system to start with.
Upon ignoring $\mathbf{A}_{i}^{(2)}$ in Eq.\ (\ref
{psiiEq}), one can solve the resulting linear differential equation for $%
\mathbf{\psi }_{i}$ and insert the solution into $\mathbf{R}$ of Eq. (\ref
{RDef}). Technically, integrating out $\mathbf{\psi }_{i}$ can be done most
conveniently in the particle's magnetization frame with the basis vectors $%
\left( \mathbf{n},\mathbf{e}_{1},\mathbf{e}_{2}\right) ,$ where $\mathbf{n}=%
\mathbf{m}/m,$ $\mathbf{e}_{1}\bot \mathbf{n,}$ and $\mathbf{e}_{2}=\left[
\mathbf{n}\times \mathbf{e}_{1}\right] $. This calculation takes into
account the time dependence $\mathbf{m}(t)$ given by Eq. (\ref{mEq0})
without the perturbation $\mathbf{R.}$ Since in a large particle spin waves
form a quasicontinuum in the $\mathbf{k}$-space, integrating over $\mathbf{k}
$ across the resonance corresponding to the SW instability
leads to the part of $\mathbf{R}$ that describes relaxation of the particle's magnetization
in Eq.\ (\ref{mEq0}). The idea of this calculation is similar to the case of
spin-lattice relaxation with replacement of phonons by internal spin waves,
however the implementation is more involved since the latter are defined in
a moving frame.

In the case of oriented anisotropy, Eq. (\ref{OADef}) or more complicated forms, the origin of the exponential SW instability leading to relaxation is the elliptic precession of spins in the spin wave. This complicates the analysis, although one can show the existence of SW instabilities \cite{kas06prl}. In the model with random anisotropy only, $D=0$, the dependence $\mathbf{m}(t)$ is a circular precession. Here the calculations simplify and lead to the equation of motion for $\mathbf{m}$ with dissipation due to the linear SW instability that can be written in the form
\begin{equation}\label{LLBSWEq}
\mathbf{\dot{m}}=\frac{1}{\hbar }\left[ \mathbf{m\times h}\right]
-m^{3/2}\left[\Gamma _{\bot }(x)\frac{\left[ \mathbf{m\times }\left[ \mathbf{m\times h}\right] \right] }{m^{2}h} - \Gamma _{\Vert }(x)\frac{\mathbf{m}}{m}\right].
\end{equation}
Here the transverse and longitudinal relaxation rates depend on the particle's magnetization orientation $x\equiv \left( \mathbf{m\cdot h}\right) /\left( mh\right)$.
\begin{eqnarray}
\Gamma _{\bot }(x) &=&\Gamma _{\bot 0}\Phi _{\bot }(x)=\frac{1}{5\pi \hbar }
\frac{D_{R}^{2}}{J}\sqrt{\frac{h}{J}}\Phi _{\bot }(x)  \label{GammaPerp} \\
\Gamma _{\Vert }(x) &=&\Gamma _{\Vert 0}\Phi _{\Vert }(x)=\frac{2}{15\pi
\hbar }\frac{D_{R}^{2}}{J}\sqrt{\frac{h}{J}}\Phi _{\Vert }(x)
\label{GammaPar}
\end{eqnarray}
with $h\ll J$ and
\begin{eqnarray}
\Phi _{\bot }(x) &=&\frac{1-x}{6}\left[ \left( 1+2x\right) ^{2}+\sqrt{2}%
(1-x)(2+x)\right]  \label{PhiPerp} \\
\Phi _{\Vert }(x) &=&\frac{\left( 1-x\right) ^{2}}{4}\left[ \left(
1+2x\right) ^{2}+\sqrt{2}\left( 1-x^{2}\right) \right] .  \label{PhiPar}
\end{eqnarray}
The applicability of our method requires $\Gamma _{\bot }, \Gamma _{\Vert } \ll \omega_H = h/\hbar$, i.e., the relaxation of the magnetization (as well as the rate of SW production) is much slower than the magnetization precession considered as unperturbed in the first approximation above.

We note that Eq.\ (\ref{LLBSWEq}) is similar to the LLB equation \cite{gar97prb, garfes04prb}. However, it is valid, in general, only for short times, when the number of excited spin waves is still small and $m\cong 1,$ so that neglecting $\mathbf{A}_{i}^{(2)}$ in Eq.\ (\ref{psiiEq}) is justified. Correspondingly, $m $ in Eq.\ (\ref{LLBSWEq}) could be replaced by 1. More consequent,
however, is to introduce the magnetization direction $\mathbf{n}$ and write $\mathbf{m=n}m.$ Then Eq. (\ref{LLBSWEq}) can be split into the two following equations
\begin{equation}
\mathbf{\dot{n}}=\frac{1}{\hbar }\left[ \mathbf{n\times h}\right] -\Gamma
_{\bot }(x)\frac{\left[ \mathbf{n\times }\left[ \mathbf{n\times h}\right] %
\right] }{h},  \label{LLBn}
\end{equation}
where $x\equiv \mathbf{n\cdot h/}h$, and
\begin{equation}
\dot{m}=-\Gamma _{\Vert }(x).  \label{LLBm}
\end{equation}

The small-$t$ behavior $m_{z}(t)=1-\Gamma _{\Vert }(-1)t$ \ following from Eq.\ (\ref{LLBm}) and shown in Fig.\ \ref{fig:DR}, agrees well with the numerical result, a small discrepancy coming from $h/J$ not being small enough. Eq.\ (\ref{LLBm}) does not describe the increase of $m$ after switching and its recovery to $m\cong 1,$ that is seen in Figs. \ref{fig:DR} and \ref{fig:Uniaxial}.

A striking feature of our result is that both $\Gamma _{\bot }(x)$ and $\Gamma _{\Vert }(x)$ vanish for $\mathbf{m}\Vert \mathbf{h}$ (i.e., for $x=1),$ while they reach their maxima at $x=-1$. Thus, initially fast relaxation slows down when the particle approaches equilibrium, see Fig.\ \ref{fig:DR}. Indeed, Eq. (\ref{LLBn}) can be rewritten as
\begin{equation}
\dot{x}=\Gamma _{\bot }(x)(1-x^{2}).  \label{xdotEq}
\end{equation}
In terms of the angular deviation from equilibrium we have
\begin{equation}\label{yDef}
y\equiv 1-\mathbf{n\cdot h/}h\equiv 1-x\ll 1
\end{equation}
upon which Eq. (\ref{xdotEq}) simplifies to $\dot{y}=-3\Gamma _{\bot 0}y^{2}$ and yields the solution
\begin{equation}
y(t)=\frac{y(0)}{1+3y(0)\Gamma _{\bot 0}t}.  \label{ySol}
\end{equation}
The long-time asymptote of this solution does not depend on the initial condition $y(0)$ and is given by $y(t)=1/(3\Gamma _{\bot 0}t).$ The full magnetization vector $\mathbf{m=n}m$ includes $m$ that follows from Eq. (\ref{LLBm}). The latter becomes $\dot{m}=-(9/4)\Gamma _{\Vert 0}y^2 = -(3/4)\Gamma _{\bot 0}y^2 = \dot{y}/4$. This yields
\begin{equation}
1-m(t)=[y(0)-y(t)]/4.
\label{mtRes}
\end{equation}
One can see that the deviation $1-$ $m(t)$ remains finite and small for $t\rightarrow\infty$, that is the consequence of slowing down the relaxation as $\mathbf{m}$ approaches equilibrium. The change of $\mathbf{m}$ due to its rotation and due to the change in its magnitude are comparable with each other. In addition, one has to remember that the equation of motion for $\mathbf{m}$ was obtained for the initial stage of relaxation only and, in particular, Eq. (\ref{LLBm}) does not describe thermalization. Still the analytical evidence of the nonexponential relaxation via internal spin waves in magnetic particles is sufficient.

In the sequel we consider both mechanisms of SW instability in magnetic particles in more physical terms and present the results of our numerical simulations.

\section{Random anisotropy and linear instability}
\label{sec-RA}
%
\begin{figure*}[ht!]
\begin{center}
\includegraphics[angle=-90,width=8cm]{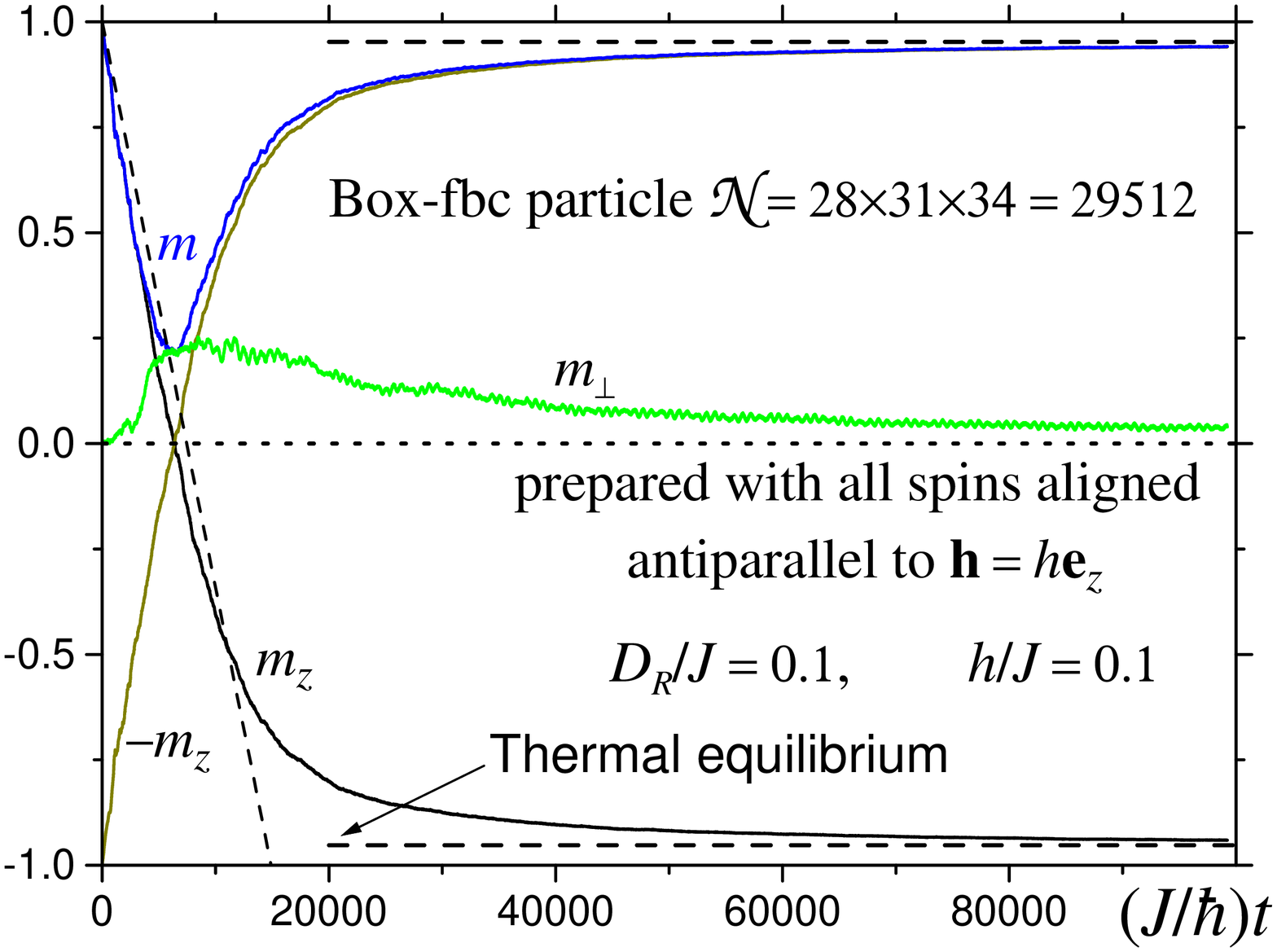}
\includegraphics[angle=-90,width=8cm]{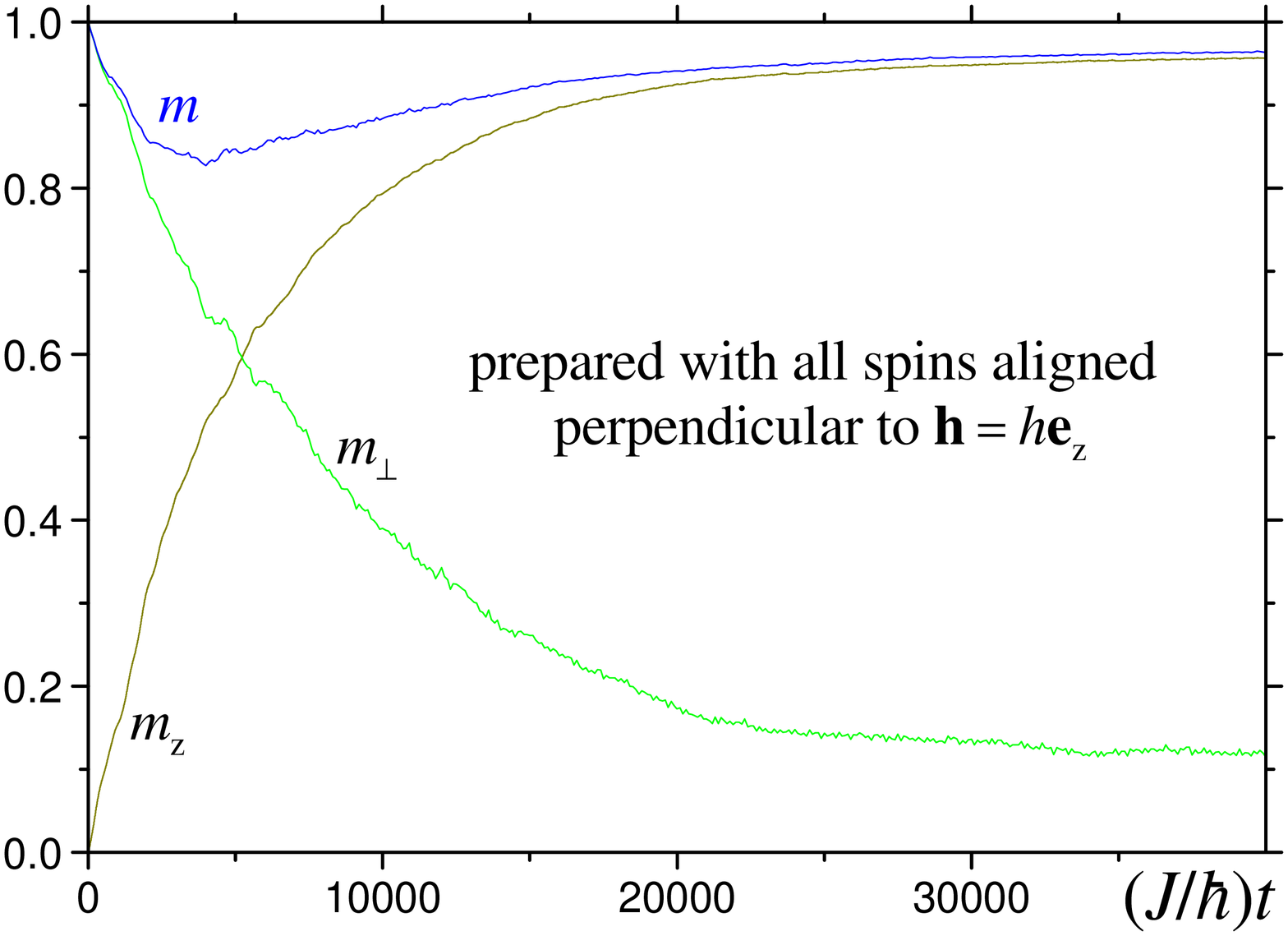}
\caption{Magnetization switching via linear spin-wave instability in a
particle with uniaxial random anisotropy. a) Particle prepared with all spins
antiparallel to the magnetic field. The small-$t$ asymptote $m_{z}(t)\cong
m(t)\cong 1-\Gamma _{\Vert }(-1)t$ of Eq.\ (\protect\ref{GammaPar}) is shown by the
dashed line. For this initial condition, switching occurs predominantly via
changing the magnetization length $m.$ b) Particle prepared with all spins
perpendicular to the magnetic field.}
\label{fig:DR}
\end{center}
\end{figure*}
%
The first particular model we study here is that with random anisotropy \cite{harplizuc73prl,chu95WS} of Eq. (\ref{RADef}).
In this case, spin waves may be generated out of a false vacuum by one-magnon processes. In particular, for $\mathbf{m}$ antiparallel to $\mathbf{h}$, spin waves energies $\varepsilon _{\mathbf{k}}$ in the particle have a negative gap $-h$. Thus a SW can be created if its energy
\begin{equation}\label{eq:energy}
\varepsilon _{\mathbf{k}}=-h+J_{0}-J_{\mathbf{k}}
\end{equation}
satisfies the resonance condition
\begin{equation}\label{ResCondSW}
\varepsilon _{\mathbf{k}} = 0.
\end{equation}
We recall that $\mathbf{k}$ in fine particles is not conserved due to the lack of translational invariance. Here $J_{\mathbf{k}}$ is the Fourier component of $J_{ij}$. For a box-shaped
particle with free boundary conditions (fbc) there are standing spin waves
with discrete wave vectors \cite{kacgar01epjb}
\begin{equation}
k_{\alpha }=\frac{\pi n_{\alpha }}{aN_{\alpha }},\quad n_{\alpha
}=0,1,\ldots ,N_{\alpha }-1,\quad \alpha =x,y,z,  \label{kfbc}
\end{equation}
where $a$ is the lattice spacing and $N_{x}N_{y}N_{z}=\mathcal{N}$. For a
particle with simple cubic (sc) lattice one has $J_{\mathbf{k}%
}=2J\sum_{\alpha }\cos (ak_{\alpha }),$ so that in the long-wave-length
limit $J_{0}-J_{\mathbf{k}}\cong J\left( ak\right) ^{2}.$ For particles
small enough, the lowest value of $J_{0}-J_{\mathbf{k}}$ for $\mathbf{k\neq 0%
}$ [i.e., $n_{\alpha }=1$ in Eq.\ (\ref{kfbc})] may exceed $h$ and thus SWs
cannot be created. This yields the absolute stability criterion for the
particle's linear size $L$
\begin{equation}
L<L_{\max }=\pi a\sqrt{J/h}  \label{LinearStabCrit}
\end{equation}
in the case of not strongly oblate or prolate particles [c.f. Eq.\ (\ref
{SingleDomainCrit})]. Under this condition, there is no relaxation and the
particle remains in its initial false-vacum state, say, antiparallel to $%
\mathbf{h}$.

For $L>L_{\max }$ the following possibilities arise. If all discrete SW modes in the particle are far from the resonance with the false vacuum and Eq. (\ref{ResCondSW}) is not satisfied, there is no relaxation as before.
If there is one mode $\mathbf{k}$ that exactly or approximately satisfies Eq.\ (\ref{ResCondSW}), this mode becomes unstable and its amplitude grows. For instance, in the sc-lattice particle with $N_{x}=N_{y}=N_{z}=5$ ($\mathcal{N}=125$) the three degenerate modes with $\mathbf{k=}(1,0,0),$ $(0,1,0)$ and $(0,0,1)$ are excited for the magnetic fields in the vicinity of $h/J=0.381966.$ For the particle initially magnetized antiparallel to the magnetic field, the magnetization, obtained numerically, is oscillating because of reversible transitions between the false-vacuum state and the state with spin waves excited. This phenomenon is similar to the probability oscillations between two resonant states in quantum mechanics. If there are several degenerate unstable $\mathbf{k}$-modes and/or the $\mathbf{k}$-modes can be converted into other spin-wave modes via nonlinear processes, the time dependence of the magnetization becomes more complex and the deviation  $1-m_{z}$ increases. Still, for small particles it does not lead to a full magnetization reversal since the spin-wave modes are discrete and it is difficult to fulfill the energy conservation laws in nonlinear SW processes.

To illustrate the linear instability in larger particles with quasi-continuous SW spectrum, we have solved Eq.\ (\ref{LLEqi}) for magnetization
switching in the random-anisotropy model.
The particle is a box with fbc, consisting of $28\times 31\times 34=29512$ atomic spins. In Fig.\ \ref{fig:DR}a we plot the results for an initial magnetization antiparallel to the field. One can see a nearly full reversal that goes linearly at small times. For this initial condition, the transverse magnetization component $m_{\bot }$ remains small.
In the middle of the switching, excited spin waves almost destroy the magnetization $m$. Then these long-wavelength SWs of high amplitude convert via nonlinear processes into all possible spin-wave modes, and the system thermalizes. Since the thermodynamics of classical spins is mainly
determined by short-wavelength modes of high energy, the energy conservation requires that the amplitudes of these SWs be small. This explains the almost full recovery of $m$ after reversal. The asymptotic disordering $1-m>0 $ (see Fig.\ \ref{fig:DR}) is due to the final temperature $T$ corresponding to the released Zeeman energy.
The numerical results in Fig.~\ref{fig:DR}b (initial magnetization perpendicular to the field) show that the magnetization relaxes mainly by rotation towards equilibrium, although $m$ decreases from saturation in the middle of the process and then recovers.
\section{Oriented anisotropy and exponential instability}
\label{sec-OA}
The second model of relaxation via internal spin waves we consider is the model with oriented uniaxial anisotropy (\ref{OADef}). In this case, the instability turns out to be exponential. The latter occurs for a general form of the single-spin part of Eq.\ (\ref{Ham}) that leads to non-circular magnetization precession, see Ref.\ \cite{kas06prl}.
For illustration, we use Eq. (\ref{OADef}) in (\ref{Ham}) with $\mathbf{h=}h\mathbf{e}_{x}$ and we choose the initial state $\mathbf{s}_{i}=\mathbf{e}_{x}.$ Linearization around the latter yields the spin-wave spectrum
\begin{equation}
\varepsilon _{\mathbf{k}}=\sqrt{\left( h+J_{0}-J_{\mathbf{k}}\right) \left(
h-2D+J_{0}-J_{\mathbf{k}}\right) }.  \label{omegakUniaxperpH}
\end{equation}

For $h>2D$ the state $\mathbf{s}_{i}=\mathbf{e}_{x}$ is the energy minimum and $\varepsilon _{\mathbf{k}}$ is real. In the interval $0<h<2D$ the state $\mathbf{s}_{i}=\mathbf{e}_{x}$ is the energy saddle point, and $\mathbf{m}$ will rotate away from this state, similarly to the case studied in Ref.~\cite{safber01prb}. SW modes in the interval $0<$ $J_{0}-J_{\mathbf{k}}<2D-h$ are unstable since $\varepsilon _{\mathbf{k}}$ is imaginary \cite{kas06prl}. For $h<0$ the state $\mathbf{s}_{i}=\mathbf{e}_{x}$ is the energy maximum, so that $\mathbf{m}$ performs small precessions around $\mathbf{e}_{x}$ in the absence of SW processes. Indeed, according to Eq.\ (\ref{omegakUniaxperpH}), the mode $\mathbf{k=0}$ is stable, so that the particle's magnetization cannot rotate away from this state. On the other hand, the modes with  $\mathbf{k\neq 0}$ in the interval $-h<$ $J_{0}-J_{\mathbf{k}}<2D-h$ are unstable. Thus this field range the only way of reversal is via excitation of internal spin waves with $\mathbf{k\neq 0}$ that strongly reduce the magnetization magnitude $m=|\mathbf{m}|$.
%

\begin{figure*}[ht!]
\begin{center}
\includegraphics[angle=-90,width=8cm]{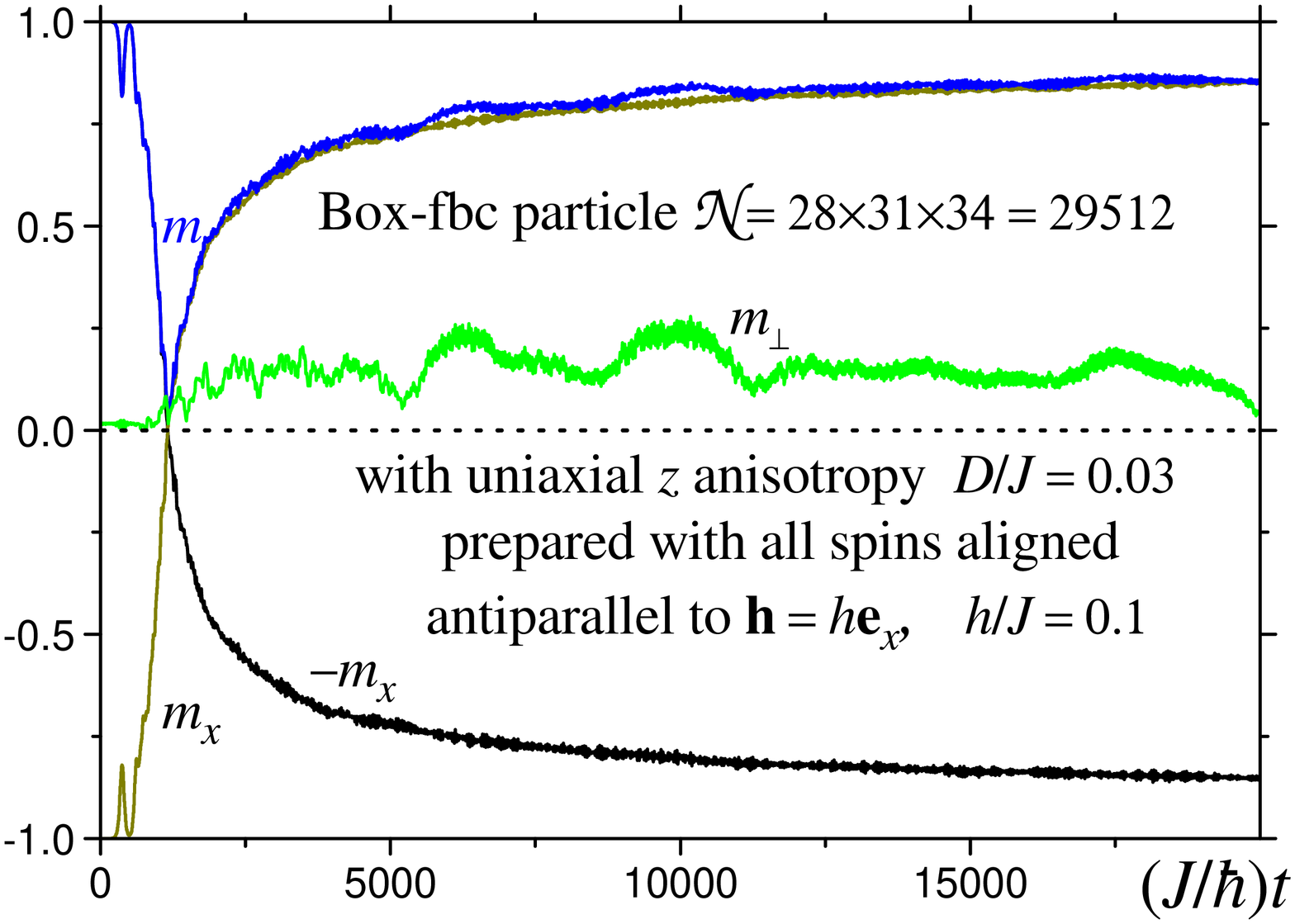}
\includegraphics[angle=-90,width=8cm]{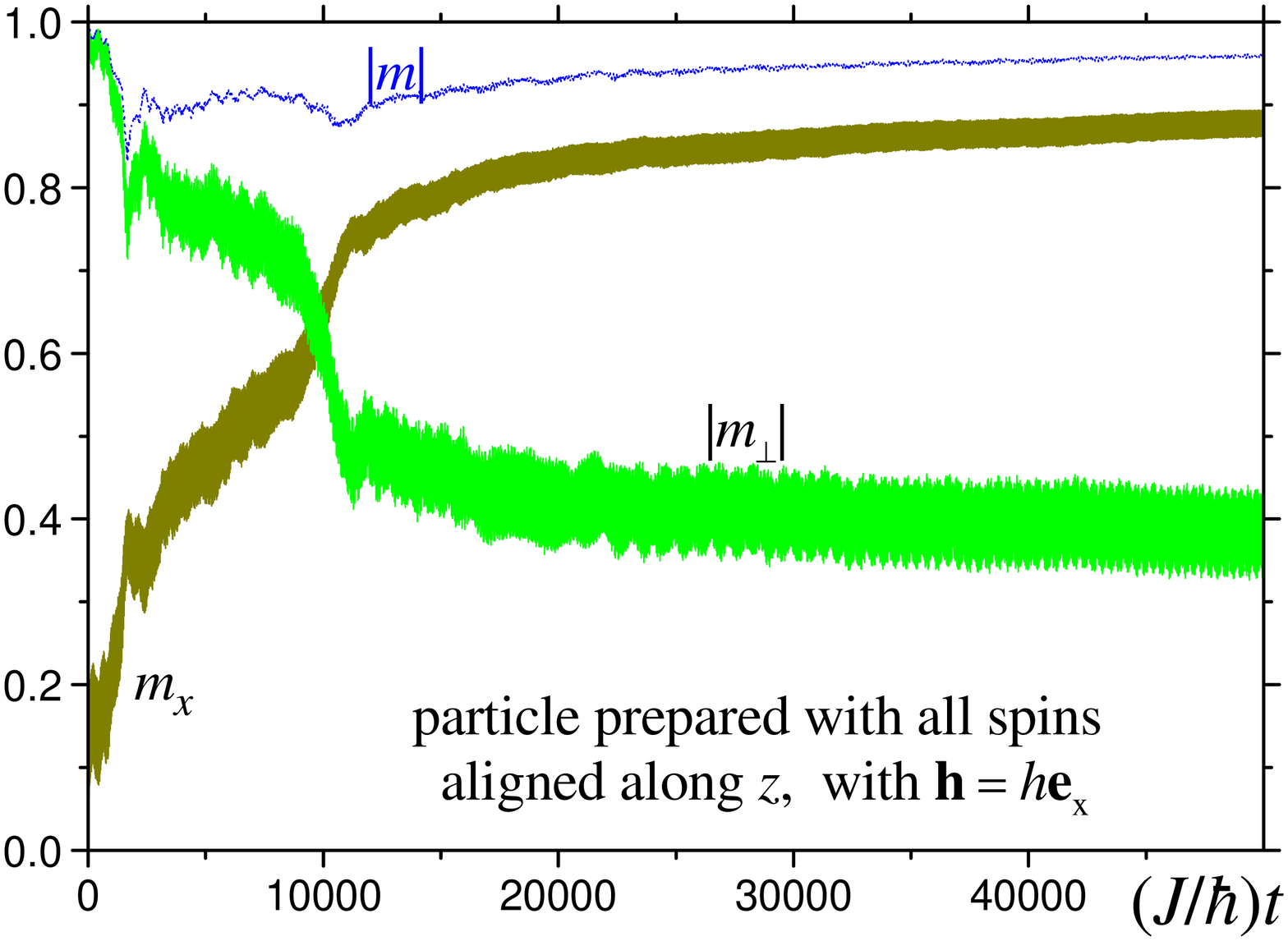}
\caption{Magnetization switching via exponential spin-wave instability in
a particle with oriented uniaxial anisotropy and transverse field. a) Particle
prepared with all spins antiparallel to the magnetic field. Again the
longitudinal relaxation is the main mechanism of switching. b) Particle
prepared with all spins perpendicular to the magnetic field.}
\label{fig:Uniaxial}
\end{center}
\end{figure*}
%

In magnetic particles, SW modes are discrete, and the instability requires
existence of SW modes inside the instability interval. The particle is \emph{%
stable} if the smallest value of $J_{0}-J_{\mathbf{k}}$ with $\mathbf{k\neq 0%
}$ (i.e., $n_{\alpha }=1)$ exceeds the right boundary of the instability
interval, i.e., $2D-h.$ This leads to the stability criterion
\begin{equation}
L_{\max }=aN_{\max }<\pi a\sqrt{\frac{J}{2D+|h|}}  \label{SingleDomainCrit}
\end{equation}
for the particle's size $L_{\max }=\max (L_{x},L_{y},L_{z}).$ Eq.\ (\ref
{SingleDomainCrit}) is similar to the single-domain criterion since its
right-hand side is the domain-wall width, if $h=0$.

Efficient magnetization reversal via the SW instability requires a
quasicontinuous SW spectrum, i.e., many SW modes in the instability
interval. For this, a \emph{strong} inequality opposite to Eq.\ (\ref
{SingleDomainCrit}) should be fulfilled, i.e., the particle's size and/or
the anisotropy $D$ should be large enough. The numerical results for the
particle of the same shape as above are shown in Fig.\ \ref{fig:Uniaxial}.
The applied magnetic field has been chosen stronger than the anisotropy
field, to make the equilibrium magnetization parallel to the field, for the
sake of comparison with Fig.\ \ref{fig:DR}. For the initial magnetization
antiparallel to the field (Fig.\ \ref{fig:Uniaxial}a), the first stage of the
relaxation is an exponential increase of $1-m$ due to the SW instability
until the first dip, followed by an incomplete recovery of $m$ and its
further decrease. The subsequent development is similar to that in Fig.\ \ref
{fig:DR}a. For the initial magnetization perpendicular to the field (Fig.\
\ref{fig:Uniaxial}b) the behavior is more complex as the components of $%
\mathbf{m}$ perform fast oscillations because of the elliptical precession.
Apart from the latter, the results at not too small times are qualitatively
similar to those in Fig.\ \ref{fig:DR}b, the main feature being a decrease and
a recovery of $m.$ The latter was observed in our simulations for other
orientations of the field with respect to the anisotropy axis.

\section{Some orders of magnitude}
Let us now estimate the magnetization reversal rates due to the internal SW
excitation. First, since Eq.\ (\ref{Ham}) is written in terms of the
normalized spins $|\mathbf{s}_{i}|=1,$ conversion to actual spins $S$
requires replacements $J\Rightarrow JS$ and $D\Rightarrow DS,$ while $h$
remains the same. For cobalt $S=3/2,$ and the above modification is
inessential for estimations. Comparison of Figs.\ \ref{fig:DR} and \ref
{fig:Uniaxial} shows that the exponential instability leads to a faster
relaxation than the linear instability. Indeed, for all materials $D\ll J$
is satisfied ($D\simeq 0.0024J$ for Co)$,$ and the instability increment in
Eq.\ (\ref{omegakUniaxperpH}) is of order $D/\hbar ,$ while Eq.\ (\ref
{GammaPerp}) contains additional small factors $D/J$ and $\sqrt{h/J}.$
However, the difference of the overall relaxation rates is not that big since
at larger times the relaxation rate should be of the order of $D^{2}$ in both cases.
Since we do not have any analytical result for the exponential scenario at
large times, we use Eq.\ (\ref{GammaPar}) for estimations. With $%
J/k_{B}\approx 10^{2}$ K for Co (assuming amorphization leading to random anisotropy), one obtains the rate $\Gamma \approx 10^{5}$ s$^{-1}$
for $h=0.001J$ that corresponds to $H\approx 0.07$ T in real units. For this
field, the precession frequency is $\omega _{H}=\left( g\mu _{B}/\hbar
\right) H\approx 10^{10}$ s$^{-1},$ i.e., precession is fast with respect to relaxation. Preparing the particle's magnetization in a desired direction requires the application of pulsed fields \cite{backetal99science,kas06prl} that are at least by an order of magnitude
larger than $H,$ during a time that is at least by an order of magnitude
shorter than $1/\omega _{H}.$ For the exponential scenario, one can expect a
similar result for $\Gamma $ without the factor $\sqrt{h/J},$ i.e., $\Gamma
\approx 3\times 10^{6}$ s$^{-1}.$ The precession frequency due to the
anisotropy field $\omega _{A}$ in Co is comparable with $\omega _{H}$ quoted
above. These estimations give an idea of the relaxation rates that remain in
the same ballpark for different kinds of the anisotropy. For instance, the
linear instability can also be driven by the surface anisotropy that is much
stronger than the volume anisotropy but acts on the surface spins only.
Numerical simulations with surface anisotropy $D_S \ll J$ give the results similar to those for the random anisotropy.
If the surface effects are so strong that another type of ordering is realized at the surface, the order in the core should be strongly distorted as well.
This will presumably lead to a very strong internal relaxation, although simulations are needed to clarify the situation.

Our numerical simulations for a multitude of different models and initial
states of the particle use exaggerated values of the anisotropy and field. Using realistic smaller anisotropy and field values leads to the necessity to work with much larger sizes, according to Eqs.\ (\ref{LinearStabCrit}) and (\ref{SingleDomainCrit}). In addition, the relaxation becomes too slow, so that much longer time interval is needed for computations. Both factors together make this impossible. Nevertheless, for particle's sizes $\mathcal{N} > 10^{4}$ the results for realistic parameters can be obtained from those for the exaggerated parameters by rescaling.

\section{Conclusion}
\label{sec-conclusion}

We have shown that exponential and linear spin-wave instabilities in large enough magnetic
particles put into a false-vacuum state (i.e., with the magnetization deviating from the equilibrium direction) can lead to complete magnetization reversal and thermalization.
Whereas the former is similar to the Suhl instability \cite{suh57,dobvic03prl}, the latter seems to have no analogue within the standard SW theory.
It should be stressed that both kinds of instabilities are described within the \emph{linear} approximation in spin-wave amplitudes, unlike the description within the standard (nonlinear) SW theory.
In both cases, the
relaxation is fast at the beginning but then slows down. These results
should be useful for investigating the dynamics of magnetoelectronic elements and, more
general, in the physics of \emph{unstable} macroscopic states.

\acknowledgments
D. G. thanks E. M. Chudnovsky for useful discussions. D. G. is a Cottrell Scholar of Research Corporation.

\end{document}